\documentclass[twocolumn]{webofc}

\usepackage[varg]{txfonts}   
\usepackage{url}
\usepackage{physics, mathtools, braket}
\usepackage{soul}
\usepackage[normalem]{ulem}   

\usepackage{caption}

\usepackage{hyperref}
\hypersetup{
    colorlinks = true,
    citecolor  = blue,
    linkcolor  = blue,
    urlcolor  = blue
}

\begin{document}

\newcommand{\Sch}{ Schr\"{o}dinger }
\newcommand{\ie}{\textit{i.e.}}
\newcommand{\eg}{\textit{e.g.}}

\newcommand{\FM}[1]{{\color{magenta} #1}}
\newcommand{\can}[1]{{\color{red}\sout{#1}}}

\definecolor{cyanobright}{rgb}{0.0,0.55,0.7}
\newcommand{\miri}[1]{\textcolor{cyanobright}{ #1}}
\newcommand{\mirix}[1]{\textcolor{cyanobright}{\sout{#1}}}

\title{Electromagnetic sum rules for $^{22}$O from coupled-cluster theory}

\author{
    \firstname{Francesco} \lastname{Marino} \inst{1}
    \fnsep \thanks{\email{frmarino@uni-mainz.de}} 
\and
    \firstname{Miriam} \lastname{El Batchy}\inst{1}
\and
    \firstname{Sonia} \lastname{Bacca}\inst{1,2}
}

\institute{
    Institut f\"{u}r Kernphysik and PRISMA+ Cluster of Excellence, Johannes Gutenberg-Universit\"{a}t Mainz, 55099 Mainz, Germany
\and
    Helmholtz-Institut Mainz, Johannes Gutenberg-Universität Mainz, D-55099 Mainz, Germany
}

\abstract{
In this contribution, we discuss recent results on the determination of the electric dipole polarizability $\alpha_D$ for the neutron-rich isotope $^{22}$O within the \textit{ab initio} framework. Our calculations utilize the Lorentz integral transform coupled-cluster (LIT-CC) approach with two different chiral potentials that incorporate two- and three-nucleon interactions. 
We compare our predictions with the  available  experimental data and find good agreement in the low-energy region.
}

\maketitle

\section{Introduction}
\label{sec: intro}
Nuclear response functions are pivotal for understanding electroweak processes on nuclei~\cite{BaccaPastore2014,Bacca2014}.
An \textit{ab initio} description of response functions up to medium-mass nuclei has been made possible by combining the coupled-cluster (CC) method~\cite{Hagen2014Review,HagenNature2015} with the Lorentz integral transform (LIT) approach~\cite{Efros_2007,Bacca2013}. 
In this work, we report on our recent progress 
in studying the response to an electric dipole field and the associated electric dipole polarizability $\alpha_D$~\cite{NeumannCosel2025}.
In particular, we focus on the neutron-rich $^{22}\rm{O}$ isotope and predict $\alpha_D$ using nuclear interactions derived from chiral effective field theory~\cite{Epelbaum2024,MACHLEIDT2024104117} that include both two- and three-nucleon forces.

\section{Theory}
\label{sec: theory}
Inclusive response functions $R(\omega)$, defined as 
\begin{align}
    \label{eq: response basic def}
    R(\omega) = \sum_{f} \mel{\Psi_{0}}{ \hat{\Theta}^{\dagger} }{ \Psi_{f}}
    \mel{\Psi_{f}}{ \hat{\Theta} }{ \Psi_{0}}
    \delta( E_{f} - E_{0} - \omega),
\end{align}
are key to describing electroweak processes involving nuclei.
The external field probing the nucleus is described by an excitation operator $\hat{\Theta}$, which connects the nuclear ground state (g.s.) $\ket{ \Psi_{0} }$ to the spectrum of both bound and continuum excited states $\ket{ \Psi_{f} }$.

Determining continuum states poses significant challenges, which can be circumvented efficiently by the LIT approach~\cite{Efros_2007,Bacca2013}.
The idea of the LIT method is to focus on the integral transform $L(\sigma,\Gamma)$ of the original energy-dependent response function $R(\omega)$, which is obtained by convoluting $R(\omega)$ with a Lorentzian kernel of centroid $\sigma$ and width $\Gamma$, namely
\begin{align}
    \label{eq: lit def}
    L(\sigma,\Gamma) = \frac{\Gamma}{\pi} \int d\omega \, \frac{ R(\omega) }{ (\sigma-\omega)^2 + \Gamma^2 } . 
\end{align}
$L(\sigma,\Gamma)$ can be determined as the norm of a bound pseudo-state, which can be solved using conventional bound-state many-body techniques~\cite{Efros_2007} (see below).
In the limit $\Gamma \to 0$, the LIT reduces to a discretized approximation of the response function in terms of a finite number of bound psuedo-states, which we denote as the discretized response function.

The LIT has been combined with the hyperspherical harmonics method~\cite{Bacca:2001kr} and the no-core shell model~\cite{Stetcu:2006dt} in light nuclei. Coupling the LIT approach with CC theory~\cite{Hagen2014Review} has been instrumental to extending the scope of \textit{ab initio} theory to the response of medium-mass nuclei~\cite{Bacca2013,Bacca2014,Miorelli2016}, including recent extensions to open-shell isotopes~\cite{Bonaiti2024,Brandherm2024,Marino2025}. CC provides an accurate parametrization of the correlated g.s.,
\begin{align}
    \label{eq: cc gs ansatz}
    \ket{\Psi_0} = e^{ \hat{T} } \ket{ \Phi_0 },
\end{align}
where $\ket{ \Phi_0 }$ is a reference state, e.g., a spherical Hartree-Fock (HF) solution in the case of closed-shell nuclei, on top of which correlations are built by the action of $e^{ \hat{T} }$, where the cluster operator $\hat{T}$ is expanded as a combination of $n$-particle-$n$-hole ($n$p-$n$h) excitation operators $\hat{T}_n$.
In the simple singles and doubles level (CCSD) approximation, $\hat{T} = \hat{T}_1 + \hat{T}_2$, with
\begin{subequations}
\begin{align}
    & \hat{T}_1 = \sum_{ai} t^{a}_{i} c_a^{\dagger} c_i\, , \\
    & \hat{T}_2 = \frac{1}{4} \sum_{abij} t^{ab}_{ij} c_a^{\dagger} c_b^{\dagger} c_j c_i\,  ,
\end{align}
\end{subequations}
where indices $i,j$ and $a,b$ denote single-particle states that are occupied (holes) and unoccupied (particles) in the reference state, respectively.
A higher accuracy can be obtained by including leading triples (3p-3h) contributions, e.g., in the CCSDT-1 truncation scheme~\cite{Hagen2014Review,Miorelli2018}.

The exponential ansatz induces a similarity (non-unitary) transformation on the operators, which reads
\begin{align}
    \Bar{H} = e^{- \hat{T}} \hat{H}_{N}  e^{\hat{T}}, \\
    \Bar{\Theta} = e^{- \hat{T}} \hat{\Theta}_{N}  e^{\hat{T}},
\end{align}
where normal-ordered operators are defined by $\hat{H}_N = \hat{H} - \mel{\Phi_0}{ \hat{H} }{\Phi_0}$ and $\hat{\Theta}_N = \hat{\Theta} - \mel{\Phi_0}{ \hat{\Theta} }{\Phi_0}$.
Since $\Bar{H}$ is non-Hermitian, the left g.s.~$\bra{\Psi_0}$ is not simply the adjoint of $\ket{\Psi_0}$, but requires solving additional equations. For spherical nuclei, $\bra{\Psi_0}$ is expressed in terms of a set of $n$h-$n$p de-excitation amplitudes $\hat{\Lambda}$~\cite{Miorelli2018,Hagen2014Review}, namely,
\begin{align}
    \bra{\Psi_0} = \bra{\Phi_0} e^{-\hat{T}} (1 + \hat{\Lambda}).
\end{align}

After having determined the $\hat{T}$ and $\hat{\Lambda}$ amplitudes, the LIT~\eqref{eq: lit def}, using the completeness relation in Eq.~\eqref{eq: response basic def}, reads~\cite{Miorelli2016,Miorelli2018}
\begin{align}
    L(\sigma,\Gamma) = \frac{\Gamma}{\pi} \innerproduct*{ \Psi_L(z^{*}) }{ \Psi_R(z) }, 
\end{align} 
where $z = \sigma + i \Gamma$ and the auxiliary states $\Psi_L(z^{*})$, $\Psi_R(z)$ are defined as the solutions to the following Schr\"{o}dinger-like equations 
\begin{align}
    & \left( \Bar{H} - z \right) \ket{ \Psi_R(z) } = \Bar{\Theta} \ket{\Phi_0}, \\
    & \bra{ \Psi_L(z^{*}) } \left( \Bar{H} - z^{*} \right) = \bra{\Phi_0} \Bar{\Theta}^{\dagger} (1 + \hat{\Lambda}),
\end{align}
with the excitation operators acting as source terms.
The auxiliary states are found by assuming an equations-of-motion (EOM) ansatz~\cite{Hagen2014Review,Miorelli2018}, in which $\ket{ \Psi_R(z) }$ is expanded, in a configuration-interaction-like way, as a vector in the space of $n$p-$n$h configurations, $\ket{ \Psi_R(z) } = \mathcal{R}(z) \ket{\Phi_0}$, where typically $\mathcal{R}(z)$ includes up to 2p-2h terms.
Using the non-symmetric Lanzos algorithm \cite{Bacca2014, Miorelli2016, Bonaiti2024}, we can compute the LIT for arbitrary $z$ by diagonalizing $\Bar{H}$ within the $n$p-$n$h space using the EOM framework~\cite{Hagen2014Review}.

Reconstructing the response function $R(\omega)$ from the LIT involves solving an ill-posed inversion problem, which has been addressed e.g.~in Refs.~\cite{Bacca2013,Efros_2007,Parnes2025}.
In contrast, moments of the response function can be computed directly from the discretized response function~\cite{Miorelli2016,Bacca2014},
\begin{align}
    \label{eq: moments def}
    m_n = \int d\omega \, \omega^{n} R(\omega) 
    = \lim_{\Gamma \to 0} \int d\sigma \, \sigma^{n} L(\sigma,\Gamma).
\end{align}
For example, $m_0$ is given by
\begin{align} 
    m_0 = \mel{\Phi_0}{ (1+\hat{\Lambda}) \Bar{\Theta}^{\dagger} \Bar{\Theta}}{\Phi_0}, 
\end{align}
and all other sum rules are defined fully by the Lanczos coefficients.

While the outlined formalism is general, from now on, we will consider the response to an electric dipole field, described by the third component of the translationally invariant dipole operator $\hat{ \mathbf{D} }$ ~\cite{Bacca2013,BaccaPastore2014}. Namely, $\hat{\Theta} = \hat{D}_{z}$, where 
\begin{align}
    \hat{ \mathbf{D} } = \sum_{i=1}^{A} ( \hat{\mathbf{r}}_i - \hat{\mathbf{R}}_{cm} ) \left( \frac{1+\hat{\tau}_{i}^{z} }{2} \right),
\end{align}
$\hat{\mathbf{r}}_i$ and $\hat{\mathbf{R}}_{cm}$ are the positions of the $i$-th particle and the center of mass, respectively, and $\hat{\tau}_i^z$ is the isospin projection.

In particular, we are interested in studying the electric dipole polarizability $\alpha_D$. 
Under the so-called unretarded dipole approximation~\cite{BaccaPastore2014,NeumannCosel2025, Golak2002}, which holds when the range of energy we are interested in is below the pion mass, $\alpha_D$ can be related to the $m_{-1}$ sum rule of the response function and to the photoabsorption cross section $\sigma_{\gamma}(\omega)$ by
\begin{align}
    \label{eq: alphaD def}
    \alpha_D = 2 \alpha \hbar c\,m_{-1} = \frac{\hbar c}{2\pi^2} \int_{0}^{+\infty} d\omega \frac{ \sigma_{\gamma}(\omega)}{ \omega^2 }
\end{align}
where $\alpha$ is the fine structure constant.

\section{Results}
\label{sec: results}

The neutron-rich $^{22}$O isotope is an interesting case study, as one of the relatively few neutron-rich nuclei for which the electromagnetic response has been accessed experimentally~\cite{NeumannCosel2025,Aumann2019}. 
In particular, photoneutron cross sections in the O chain were extracted in Ref.~\cite{O22response} from the measured electromagnetic excitation cross sections using heavy ions~\cite{Aumann2019}.

We investigate dipole excitations of the $^{22}\rm{O}$ nucleus using the LIT-CC method employing two chiral Hamiltonians containing two- and three-nucleon forces:
$\rm{NNLO_{sat}}$~\cite{NNLOsat} and $\Delta\rm{NNLO_{GO}(394)}$~\cite{DeltaGo2020}. Both interactions accurately describe bulk nuclear observables up to medium-mass nuclei.
As in Ref.~\cite{Miorelli2018}, we perform calculations varying the truncation level in the g.s.~between CCSD and CCSDT-1. The EOM equations for the auxiliary states are solved by including contributions up to 2p-2h excitations ~\cite{Marino2025}. The transition operator $\Bar{\Theta}$ is approximated using the $\hat{T}_1$ and $\hat{T}_2$ amplitudes from the g.s. calculation~\cite{Miorelli2018}.
We denote these two truncation schemes as "D" and "T-1", respectively. 

In all computations, the spherical HF state is built from a model space consisting of up to $N_{max}$+1 =15 major harmonic oscillator (HO) shells. 
To determine the optimal HO frequencies, we have determined $\alpha_D$ for different values of $N_{\text{max}}$ and frequency $\hbar\omega$. The best frequency is chosen as the one for which results for successive model-space truncation (namely, $N_{max}=12$ and $N_{max}=14$) are closest~\cite{Marino2025}.
We find them to be $\hbar\omega = 12\,\rm{MeV}$ for $\Delta\rm{NNLO_{GO}(394)}$ and $\hbar\omega = 14\,\rm{MeV}$ for $\rm{NNLO_{sat}}$.

\begin{figure}[h!]
    \centering
    \includegraphics[width=\columnwidth]{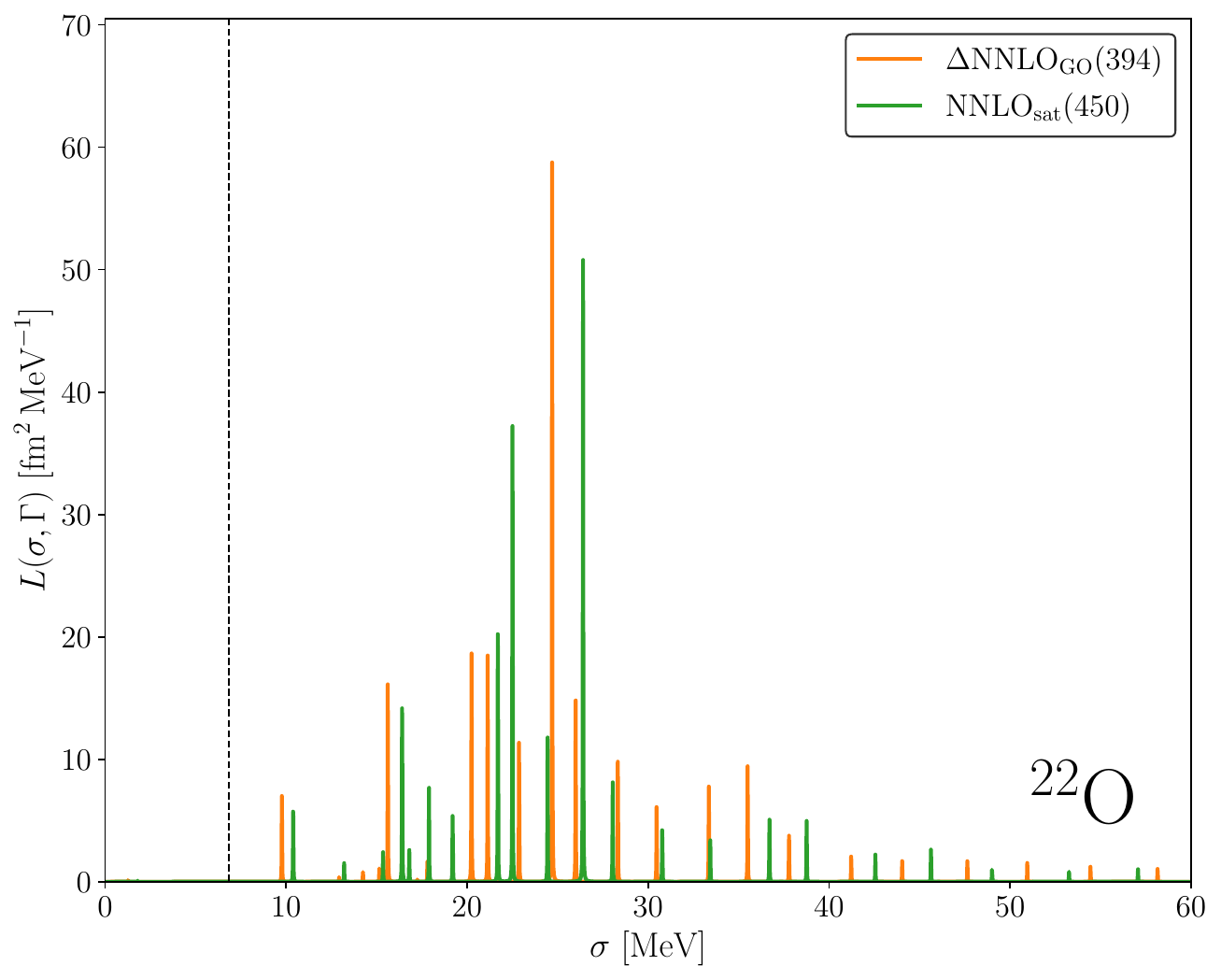}
    \caption{Discretized response function $L(\sigma,\Gamma)$ for $\Gamma=10^{-2}\,\rm{MeV}$, obtained with the T-1 truncation with two interactions (see legend).
    The experimental neutron separation energy is indicated by a black dashed line.
    }
    \label{fig: LITs}
\end{figure}

In Fig.~\ref{fig: LITs}, we show the discretized response functions, $L(\sigma,\Gamma)$ for $\Gamma=10^{-2}\,\rm{MeV}$, obtained with the two interaction employing the T-1 truncation scheme.
Also reported in the figure is the experimental neutron separation energy of 6.9 MeV (dashed line)
\footnote{Separation energies are taken from \url{https://atom.kaeri.re.kr/nuchart/}.}.
Both calculations predict dipole strength at about 10 MeV, close to the threshold for the neutron decay channel, which is consistent with the low-energy mode observed in the data~\cite{O22response,Bacca2014}.
A concentration of strength is noticed at about 20-25 MeV corresponding to the giant dipole resonance (GDR).

\begin{figure}[h!]
    \centering
    \includegraphics[width=\columnwidth]{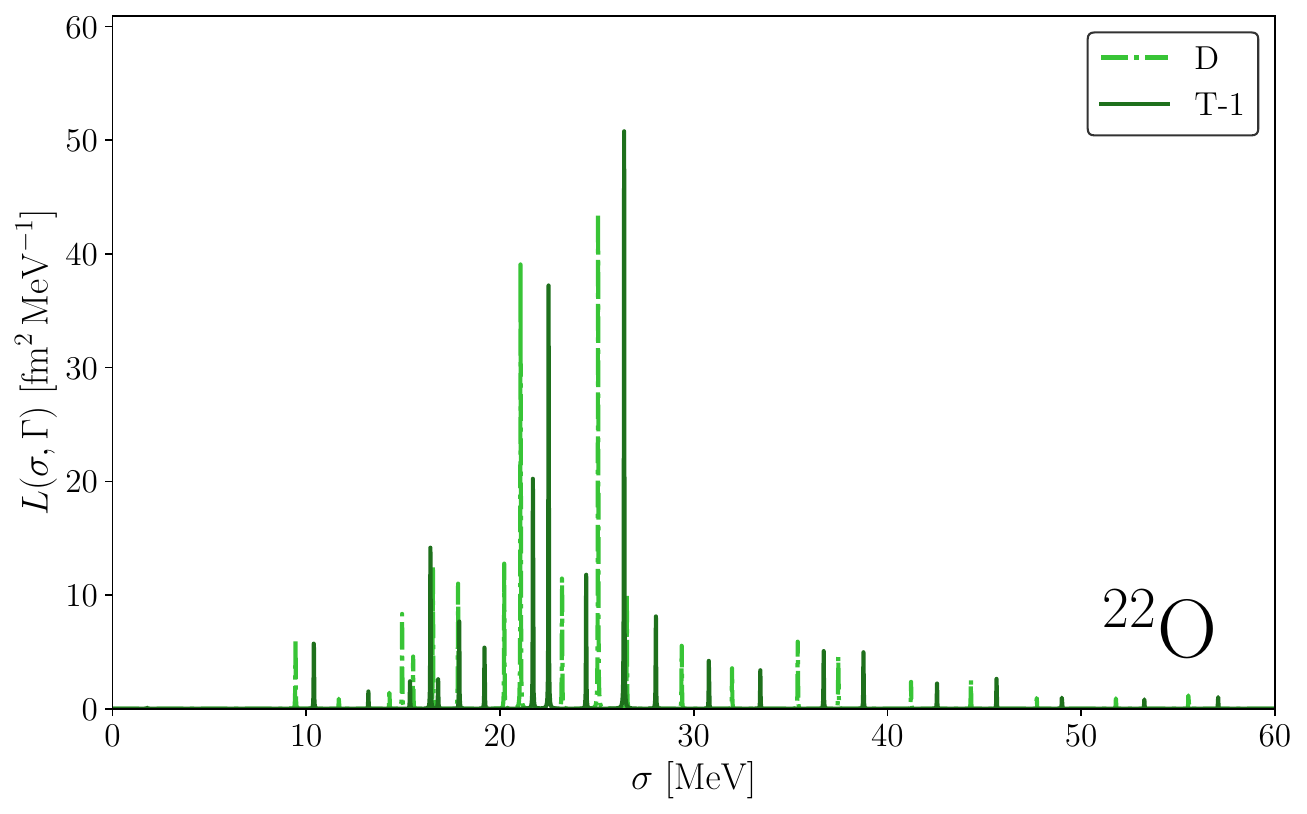}
    \caption{Discretized LIT for $\Gamma = 10^{-2}$ obtained for the $\rm{NNLO_{sat}}(450)$ interaction varying the truncation in the g.s. between CCSD ("D" in the legend) and CCSDT-1 ("T-1"). 
    }
    \label{fig: LIT different schemes}
\end{figure}

To assess the dependence of our results on the g.s. truncation scheme, in Fig. \ref{fig: LIT different schemes} we compare the discretized response functions obtained at the D and T-1 levels using the $\rm{NNLO_{sat}}(450)$ interaction.
The distribution of strength in the GDR region is roughly similar in the two cases, although peaks in the T-1 calculation are slightly shifted to higher energies.
This behavior is consistent with previous findings (see e.g. \cite{Bonaiti2024,Miorelli2018}) and can be linked to a reduction of the dipole sum rules, which is typically observed when including triple corrections. 

\begin{figure}
    \centering
    \includegraphics[width=\columnwidth]{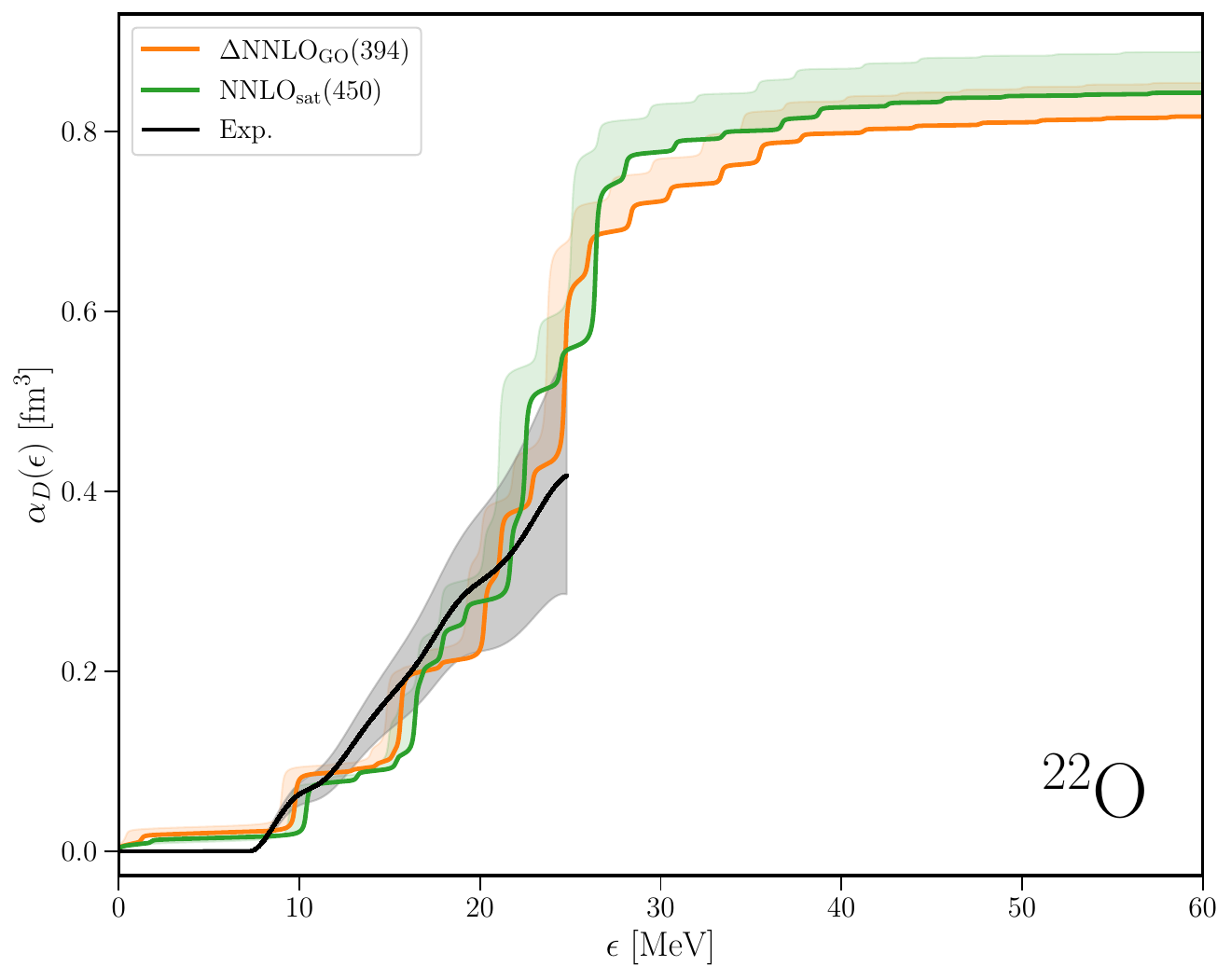}
    \caption{
    $\alpha_D$ running sum as a function of the maximum energy $\epsilon$. 
    Theoretical predictions at the T-1 level (see text) are shown as solid lines for two chiral interactions.
    The shaded bands indicate the difference between calculations at the T-1 and D truncation levels.
    The sum rule associated with the experimental data of Ref.~\cite{O22response} is depicted as a black line, with the shaded gray area accounting for the uncertainties propagated from the measured cross sections to $\alpha_D(\epsilon)$.
    }
    \label{fig: running_sum_O22}
\end{figure}

In Fig.~\ref{fig: running_sum_O22}, running sums $\alpha_D(\epsilon)$, obtained by integrating the relation Eq.~\eqref{eq: alphaD def} up to a finite energy $\epsilon$, 
\begin{align}
    \alpha_D(\epsilon) = 2\alpha \hbar c \int_{0}^{\epsilon} d\sigma 
    \frac{L(\sigma,\Gamma')}{\sigma}\,,
\end{align}
are then considered, with $\Gamma' = 10^{-4}$ MeV.
The theoretical results at the T-1 level associated with the LITs of Fig.~\ref{fig: LITs} are shown as solid lines, with shaded bands indicating the difference between T-1 and D results.
However, a direct comparison with the experimental running sum (black line) requires additional caveats.
While the photoneutron cross section approximates well the total photoabsorption cross section in heavy nuclei such as $^{208}$Pb~\cite{NeumannCosel2025}, in lighter nuclei it only represents a lower bound to the latter due to the unresolved decay channels~\cite{Goriely2020,NeumannCosel2025}, e.g., proton emissions above the proton separation energy ($23.2~\rm{MeV}$ for $^{22}\rm{O}$). 
Moreover, in the case of the unstable $^{22}\rm{O}$ isotope, the cross section was measured only up to $24.9 \, \rm{MeV}$. Also, few data points are available and are affected by large uncertainties, thus requiring special care in both evaluating and interpreting the resulting polarizability.
We have interpolated the experimental sum rule with a cubic spline before integrating it. 
The gray band denotes a conservative estimate of the errors on $\alpha_D(\epsilon)$, which we have obtained by integrating the upper and lower bounds of the experimental cross sections.

To consistently compare the different results, we report in  Tab.~\ref{tab: alpha running} the values of $\alpha_D(\epsilon)$ in the restricted energy range up to $24.9 \, \rm{MeV}$, in addition to LIT-CC results for $\alpha_D$ in the limit of large maximum energy [$\alpha_{D}(\epsilon\to\infty)$].
The quoted uncertainties on the LIT-CC predictions are estimated as half of the difference between the D and T-1 truncations~\cite{Bonaiti2024} and reflect the many-body uncertainty, which dominates over the model-space-dependence error~\cite{Marino2025}.

In Fig.~\ref{fig: running_sum_O22}, we notice a good agreement between theory and experiment for the dipole polarizabilities in the low-energy part of the spectrum ($\epsilon < 24.9\,\rm{MeV}$).
The LIT-CC for $\epsilon \to \infty$ becomes larger and exceeds the experimental value at $\epsilon = 24.9~\rm{MeV}$, most likely because the latter misses charged-particle emission components, which contribute to the GDR~\cite{Goriely2020,Aumann2019}.
It is interesting to observe that the predictions for the total $\alpha_D$'s with the $\Delta \rm{ NNLO_{GO}(394) }$ and $\rm{ NNLO_{sat} }$ potentials are consistent with previous calculations~\cite{Miorelli2018,Bonaiti2024} and agree well with each other. In this specific case, the impact of the interaction dependence is small and comparable to the many-body error.

In Tab.~\ref{tab: alpha running}, we also report $\alpha_D$  for the Entem-Machleidt interaction EM(500)~\cite{EntemMachleidt} (without three-nucleon forces). 
These values have been estimated by using the Lanczos coefficients obtained in Ref.~\cite{Bacca2014} at the CCSD level.
The results highlight that the inclusion of three-nucleon forces and of triples correlations, achieved in this work, improves the description of the low-energy strength distribution. 
Finally, the extrapolated $\alpha_D (\epsilon \to \infty)$ obtained for EM(500) is slightly larger than our results, indicating that the inclusion of three-nucleon forces and triples also leads to a redistribution of the strength at higher energies.

\begin{table}[h!]
    \centering
    \caption[textwidth = 0.8\textwidth]{Experimental running sum rule $\alpha_D(\epsilon)$ evaluated integrating the cross section up to the maximum available excitation energy, $\epsilon=24.9\,\rm{MeV}$, compared to theoretical estimates obtained by integrating the LIT up to the maximum energy probed experimentally and up to large energies ($\epsilon \to \infty$). 
    Uncertainties (see text) on both the experiment and the theory predictions for $\Delta \rm{ NNLO_{GO}(394) }$ and $\rm{ NNLO_{sat} }$ are shown in parenthesis.
    Predictions for the EM(500) interaction~\cite{EntemMachleidt} are extracted from the results of Ref.~\cite{Bacca2014}. 
    All values are in $\rm{fm}^{3}$. 
    }
    \begin{tabular}{ p{2.5cm} p{3cm} p{2cm} }
        & $\alpha_{D}(\epsilon=24.9\,\rm{MeV})$ & $\alpha_{D}(\epsilon\to\infty)$ \\
        \noalign{\vskip 1.mm} 
        \hline
        \noalign{\vskip 1.5mm} 
         Exp. & 0.41 (13) & \\
         $\Delta \rm{ NNLO_{GO}(394) }$ & 0.61 (4) & 0.83 (2)\\
         $\rm{ NNLO_{sat} }(450)$ & 0.56 (4) & 0.86 (3) \\
         EM(500) & 0.69 & 0.91
    \end{tabular}
    \label{tab: alpha running}
\end{table}

\section{Conclusions and perspectives}
In this work, we have presented an \textit{ab initio} study of the electric dipole polarizability $\alpha_D$ of the neutron-rich isotope $^{22}\rm{O}$. The calculations were performed using the LIT-CC method employing two different chiral interactions, $\rm{NNLO_{sat}}(450)$ and $\Delta\rm{NNLO_{GO}(394)}$, both including two- and three-nucleon forces.

Our results show that both interactions predict dipole strength near the neutron emission threshold, consistent with the low-energy mode observed experimentally~\cite{O22response}.
Our calculations are also able to describe the giant dipole resonance, located around $20-25$~MeV.
The calculated $\alpha_D$ shows a mild dependence on the adopted truncation scheme, thus suggesting that predictions are robust.

A comparison to the experimental data of Ref.~\cite{O22response} is somewhat delicate, since measurements are available only up to 24.9 MeV and are unable to resolve the GDR region completely.
However, when confronting $\alpha_{D}$ over the same energy range, we find a satisfying agreement of our results with the experiment.
This hints at the ability of LIT-CC to correctly model the dipole response in $^{22}\rm{O}$.

In conclusion, the LIT-CC method represents a robust framework for studying electromagnetic sum rules also in neutron-rich isotopes.
Ongoing work is now devoted to reconstructing the response function $R(\omega)$ via inversion of the LIT, thus allowing for a comparison with experimental cross sections in both closed- and open-shell~\cite{Bonaiti2024,Marino2025} nuclei.

\section{Acknowledgements}
We thank Gaute Hagen for sharing the nuclear coupled cluster code developed at Oak Ridge (NUCCOR).
This work was supported in part by the Bundesministerium für Forschung, Technologie und Raumfahrt (BMFTR, Germany) under contract number 05P2024 (ErUM-FSP T07) and in part by the Deutsche Forschungsgemeinschaft (DFG, German Research Foundation) through
Project-ID 279384907 – SFB 1245 and through the Cluster of Excellence “Precision Physics, Fundamental Interactions, and Structure of Matter” (PRISMA+ EXC
2118/1, Project ID 39083149).
Calculations were performed using the supercomputer Mogon at Johannes
Gutenberg Universit\"at Mainz.

\bibliography{bibliography.bib} 

\end{document}